\documentclass[amsmath,amssymb,nofootinbib]{revtex4}

\usepackage{latexsym,comment}
\usepackage{amssymb}
\usepackage{amsfonts}
\usepackage{amsmath,color}
\usepackage[dvips]{graphicx}
\usepackage{bm}

\def\mb#1{\mathbf{#1}}

\def\ber{\begin{eqnarray}}
\def\eer{\end{eqnarray}}
\def\beq{\begin{equation}}
\def\eeq{\end{equation}}

\def\rmd{{\rm d}}

\def\dal{\mathop{\rlap{\hbox{$\sqcap$}}\sqcup}\nolimits}   
\newcommand{\ppar}[2]{\frac{\partial #1}{\partial #2}}

\begin{document}

\title{Gravito-electromagnetic Effects of Massive Rings}

\author{Matteo Luca Ruggiero}
\email{matteo.ruggiero@polito.it}
 \affiliation{DISAT, Politecnico di Torino, Corso Duca degli Abruzzi 24, Torino, Italy\\
 INFN, Sezione di Torino, Via Pietro Giuria 1, Torino, Italy}

\date{\today}

\begin{abstract}
The Einstein field equations in linear post-Newtonian approximation  can be written in analogy with electromagnetism, in the so-called gravito-electromagnetic formalism. We use this analogy to study the  gravitational field of a  massive ring: in particular, we consider a continuous mass distribution on  Keplerian orbit around a central body, and we work out the gravitational field generated by this mass distribution in the intermediate zone between the central body and the ring, focusing on the gravito-magnetic component that originates from the rotation of the ring. In doing so, we generalize  and complement some previous results that focused on the purely Newtonian effects of the ring (thus neglecting its rotation) or that were applied to the case of rotating spherical shells.
Eventually, we study in some simple cases the effect of the the rotation of the ring, and suggest that, in principle, this approach could be used to infer information about the angular momentum of the ring.
\end{abstract}

\maketitle

\section{Introduction}\label{sec:intro}


 {The existence of the so-called gravitomagnetic (GM) effects, somehow guessed by Einstein himself \cite{einstein1,einstein2,besso}, is deeply rooted in the Equivalence Principle. Indeed, when arbitrary, time varying and nonuniform accelerated frame are considered, it is straightforward to postulate the existence of gravitational effects analogous to the well known effects arising in non-inertial frames (centrifugal, Coriolis and  angular acceleration effects). They were calculated by Einstein himself \cite{einsteinthirring}, and others \cite{lensethirring,thirring1,thirring,thirring2,mas93,iorio2002,binind,iorio2012}. Historical reconstructions and analyses are in \cite{pfister1,pfister2}.\\
\indent Actually, in General Relativity (GR) a GM field is  generated by mass currents, in close analogy with classical electromagnetism: more in general, 
the field equations of GR, in linear post-newtonian approximation, can be written in form of Maxwell equations for the gravito-electromagnetic (GEM) fields \cite{gem,mashhoon01}, \cite{mashhoon03}.}



 {As far as the so-called Lense-Thirring effect \cite{lensethirring} is concerned, that is the precessions of the node and the periapsis of a satellite which orbits a central spinning mass, according to \cite{pfister1}, it should be named more properly as Einstein-Thirring-Lense effect. Indeed, in 1917, Einstein \cite{einsteinthirring} wrote to Thirring that he calculated the Coriolis-type field of the rotating Earth and Sun, and its influence on the orbital elements of planets and satellites. In \cite{iorio2012}, the Lense-Thirring precessionsÊ were calculated for an arbitrary direction of the angular momentum of the source.Ê About the experimental efforts to measure it, they are somehow disputed: they are theÊ ongoing LAGEOS tests with the Earth \cite{Ciufolini:2004rq,ciufolini2010}, the MGS tests with Mars \cite{iorio2006,iorio2010a} and the ongoing tests with Sun and planets \cite{iorio2012a}.  Overviews are in \cite{ciufolini2007,Iorio:2010rk,iorio2003a,renzetti2013a}. The LARES mission \cite{LARES} has been launched in February 2012 for measuring the Lense-Thirring effect, and is now under way and gathering data; it is uncertain if it will be finally able to reach is ambitious goal \cite{iorio2009x,renzetti2013b,renzetti2012,ciufolini2010x}.}

 {The GM effect measured by the GP-BÊ \cite{GP.B} is a different one, that is the precession of orbiting gyroscopes, known also as Pugh-Schiff effect \cite{pugh,schiff}.} 

 {The peculiarities of the space-time around a spinning mass are revealed by the GM clock effect, that is the difference in the proper  periods of standard clocks in prograde and retrograde circular orbits around a rotating mass; however, despite several studies and proposal to measure it, this effect has not been measured yet \cite{clock1,clock2,clock3,clock4,clock5,clock6,clock7,clock8,clock9,clock10,clock11,clock12}.}

 {The possibilty of testing GM effects in a terrestrial laboratory has been explored by various author in the past (see e.g. \cite{braginsky1,braginsky2,cerdonio,mach1,camacho,iorio2003y,pascual,stedman,iorio2006w}); more recently, the proposal of testing GM effects in the Earth by means of an array of ring lasers has been considered \cite{GINGER11}, and is nowadays under development \cite{GINGER14}.}

In this paper we exploit the GEM analogy to write the field of a thin massive ring. To be more specific, we consider a continuous mass distribution on a Keplerian orbit around a central body, and we work out the gravitational field generated by this mass distribution:  {in particular, we focus on the relativistic GM effects, since the Newtonian ones have been already thoroughly studied elsewhere \cite{Iorio:2012tp}.} Actually, massive rings are ubiquitous and important in astrophysics.   Just to give some examples, as suggested by the exhaustive Introduction in \cite{Iorio:2012tp}, the giants planets in the Solar System are surrounded by rings, there is the possibility that circumsolar massive rings do exist, the minor asteroids between Mars and Jupiter con be modeled as a continuous ring and there is evidence that similar structures are present in extra solar systems,  {such as the exoplanet J1407b \cite{exo}}. Moreover, rings are present in different astrophysical situations, such as around supermassive black holes, and ring-like structures are important in galaxies formation and in the study of the interactions between galaxies. Eventually, debris of human artifacts  form annular structures around the Earth. In all these cases, it is important to evaluate the impact of the rings on the dynamics of celestial objects.  Similar situations are considered, in a full GR framework, focusing on toroid mass configurations around black holes \cite{nishida94}, three-dimensional (ring-like) distributions of matter in galaxies\cite{vogt05}, self-gravitating and rotating matter around black holes \cite{khanna92} (see, again, \cite{Iorio:2012tp}, for a comprehensive references list). 

 {GM effects are relevant also in gravitational lensing \cite{sereno}; in particular,} the GM time delay due to the propagation of light rays in presence of rotating sources could be interesting for the determination of the properties of astrophysical objects. For instance, the GM time delay in the field of a rotating source and inside a \textit{spinning shell} was studied in  \cite{ciufolens}, where it is suggested that, in principle, it can be used to estimate the angular momentum of the rotating body.

As we have seen, ring configurations are common
in astrophysics, so it could be interesting to obtain simple expressions of the gravitational field inside a \textit{rotating ring:}  {we already mentioned that} the Newtonian (i.e. gravito-electric, GE) part of the field and its effects on the orbits of test masses was studied in \cite{Iorio:2012tp}; as for the GM  part, a simple approach was considered in \cite{ashbygem}, where the case of a rotating circular ring was studied. Here we generalize and complement these approaches and calculate the gravitational field of a rotating ring (which is, in general, elliptical) as a perturbation of the background field determined by the central body; in our  model, we assume that the ring is thin and that its matter distribution has constant density.
In particular, we focus on the GM field  and discuss its impact on some observational tests.

In the first part of the paper (Section \ref{sec:GEM}) we review the GEM formalism, to give the reader a self-consistent introduction to this approach; then,  in Section \ref{sec:GEMf} we study the GE and the GM field of a rotating ring. Eventually, in Section \ref{sec:conc} we discuss the impact of the ring field on some observational tests, while the conclusions are  drawn in Section \ref{sec:cconc}.

\section{GEM in a Nutshell}\label{sec:GEM}

The Einstein field equations in linear post-Newtonian approximation can be written in analogy with electromagnetism. To this end, we  consider the space-time around  localized sources, which are allowed to rotate slowly: hence, we can write the space-time metric in the \textit{weak-field and slow-motion approximation} in the form\footnote{Greek indices run to 0 to 3, while Latin indices run from 1 to 3; 
 bold face letters like ${\mathbf{x}}$ refer to space vectors.} $g_{\mu\nu}=\eta_{\mu\nu}+h_{\mu\nu}$, in terms of the Minkowski metric tensor $\eta_{\mu\nu}$ and the first order perturbations $h_{\mu\nu}$, that are called \textit{gravitational potentials}. In what follows we  refer to the convention used in \cite{mashh1,mashh2,mashh3,mas93,mas99,binind} for the defintion of the GEM fields\footnote{Even though there  are other conventions, see e.g. \cite{ciufoliniwheeler,ruffi}.}; in addition, the space-time signature is assumed to be $+2$. It is possible to show that, if we perform the coordinates transformation  $x^\mu=(ct,{\mathbf x})$, $x^\mu \mapsto x^\mu-\epsilon^\mu$ (gauge transformation), the gravitational potentials transform  as $h_{\mu\nu}\mapsto h_{\mu\nu}+\epsilon_{\mu, \nu}+\epsilon_{\nu ,\mu}$, hence they are gauge-dependent. If we introduce the potentials $\bar h_{\mu\nu}=h_{\mu\nu}-\frac12 h \eta_{\mu\nu}$ with $h={\rm tr}(h_{\mu\nu})$, and impose the transverse gauge condition $\bar h^{\mu\nu}{}_{,\nu}=0$, the gravitational field equations take the form
\beq
\dal \bar h_{\mu\nu}=-\frac{16\pi G}{c^4}T_{\mu\nu}\ . \label{eq:fieldgem1}
\eeq

The analogy with the equations for the electromagnetic (EM)  four-potential is manifest. Indeed, in terms of the EM charge density $\rho_{EM}$ and current $j_{EM}^{i}$, we write the four-current $j_{EM}^{\mu}=\left(c\rho_{EM}, \mb j_{EM}\right)$; similarly, in terms of the electric $\Phi_{EM}$ and magnetic $A_{EM}^{i}$ potentials, we write the four-potential $A_{EM}^{\mu}=\left(\Phi_{EM}, \mb A_{EM} \right)$. Then, on using  the \textit{Lorentz gauge} condition $\displaystyle \frac{1}{c} \ppar{\Phi_{EM}}{t}+\bm \nabla \cdot \mb A_{EM} \equiv \partial_{\mu}A_{EM}^{\mu}=0$, the  equations for $A_{EM}^{\mu}$ are (see e.g. \cite{jackson})
\beq
\dal A_{EM}^{\mu}=\frac{4\pi}{c}j_{EM}^{\mu} \label{eq:boxAmu}
\eeq
If we neglect the wave-like solution of the homogeneous equations,  $A^{\mu}_{EM}$ is expressed in terms of the \textit{retarded potentials}
\beq
A_{EM}^{\mu}(ct,\mb x)=\frac 1 c \int_{V} \frac{j^{\mu}_{EM}(ct-|{\mathbf x}-{\mathbf X}|, {\mathbf X})}{|{\mathbf x}-{\mathbf X}|}\rmd V, \label{eq:solem1}
\eeq
where integration is carried out on the domain $V$ containing the charges. In particular, we get the following expressions for the electric and magnetic potentials
\beq
\Phi_{EM}(ct,\mb x)=\int_{V} \frac{\rho_{EM}(ct-|{\mathbf x}-{\mathbf X}|, {\mathbf X})}{|{\mathbf x}-{\mathbf X}|}\rmd V, \quad A_{EM}^{i}(ct,\mb x)= \frac 1 c \int_{V} \frac{j_{EM}^{i}(ct-|{\mathbf x}-{\mathbf X}|, {\mathbf X})}{|{\mathbf x}-{\mathbf X}|}\rmd V. \label{eq:solem1phiAi}
\eeq
Now, if we come back to the linearized gravitational field equations (\ref{eq:fieldgem1}) and 
neglect the gravitational waves solution of the  homogeneous  equations, in  analogy with (\ref{eq:solem1}) we can write the general solution in the form
\beq
 {\bar h}_{\mu\nu}(ct,\mb x)=\frac{4G}{c^4}\int_{V} \frac{T_{\mu\nu}(ct-|{\mathbf x}-{\mathbf X}|, {\mathbf X})}{|{\mathbf x}-{\mathbf X}|}\rmd V , \label{eq:solgem1}
\eeq
where integration is extended to the domain  $V$ containing the masses. We are interested in the weak-field and slow-motion solutions, that is in the linearized GEM approach\footnote{The fact that the Einstein field equations can be written in analogy with electromagnetism in full theory, without approximation, is well known: see e.g. \cite{LyndenBell:1996xj} and \cite{LL}, \S 95.}: as a consequence, we may neglect in the metric tensor terms that are $O(c^{-4})$. We may set $T^{00}=\rho_{G  } c^2$ and $T^{0i}=cj_{G  }^i$, in terms of the mass density $\rho_{ G }$ and mass current $j_{ G }^{i}$ of the sources, so that   $j_{G  }^\mu=\left(c\rho_{ G },{\mathbf j_{  G}}\right)$ is the mass-current four vector. Hence, from (\ref{eq:solgem1}) we get the following non null components of the tensor $\bar h_{\mu\nu}$:
\beq
{\bar h}_{00}(ct,\mb x)=\frac{4G}{c^{2}}\int_{V}  \frac{\rho_{G}(ct-|{\mathbf x}-{\mathbf X}|, {\mathbf X})}{|{\mathbf x}-{\mathbf X}|}\rmd V,   \quad
{\bar h}_{0i}(ct,\mb x)=-\frac{4G}{c^{3}}\int_{V} \frac{j_{G}^{i}(ct-|{\mathbf x}-{\mathbf X}|, {\mathbf X})}{|{\mathbf x}-{\mathbf X}|}\rmd V. \label{eq:solgemh0i}
\eeq
So, if we define the gravito-electric  (GE) $\Phi_{ G }$  and gravito-magnetic (GM) $A_{G\, i}$ potentials by
\beq
\bar h_{00} \doteq 4\frac{\Phi_{ G }}{c^{2}}, \quad \bar h_{0i}=-2 \frac{A_{G\, i}}{c^{2}} \label{eq:defphiAigem}
\eeq
we get the following expressions in terms of the sources of the gravitational field
\beq
\Phi_{G} (ct,\mb x)={G}\int_{V}\frac{\rho_{G}(ct-|{\mathbf x}-{\mathbf X}|, {\mathbf X})}{|{\mathbf x}-{\mathbf X}|}\rmd V, \quad 
A_{G \, i}(ct,\mb x)=\frac{2G}{c}\int_{V} \frac{j_{G}^{i}(ct-|{\mathbf x}-{\mathbf X}|, {\mathbf X})}{|{\mathbf x}-{\mathbf X}|}\rmd V.\label{eq:solgemAi1}
\eeq

A comparison with the corresponding equations defining the electromagnetic potentials (\ref{eq:solem1phiAi}) shows that, while the definition of the gravito-electric potential in terms of the mass density is analogous to the definition of the electric potential in terms of the charge density, the definition of the gravito-magnetic potential in terms of the mass current density is different by a factor 2 with respect to the corresponding definition of the magnetic potential in terms of the charge current density.  Eventually, we get the following expression for the space-time metric
\beq
\mathrm{d} s^2= -c^2 \left(1-2\frac{\Phi_{G}}{c^2}\right)\rmd t^2 -\frac4c ({\mathbf A_{G}}\cdot \rmd {\mathbf x})\rmd t +
 \left(1+2\frac{\Phi_{G}}{c^2}\right)\Delta_{ij}\rmd x^i \rmd x^j\, \label{eq:weakfieldmetric1}
\eeq

\noindent If we define the gravito-electric $\mb E_{G}$ and gravito-magnetic $\mb B_{G}$ fields  by
\beq
\mb E_{G}= -\frac{1}{2c} \ppar{\mb A_{G}}{t}-\bm \nabla \Phi, \quad  \mb B_{G}= \bm \nabla \wedge \mb A_{G} \label{eq:solgemEB1}
\eeq
they fulfill the following Maxwell-like equations
\begin{eqnarray}
\bm \nabla \cdot \mb E_{G}&=& 4\pi G \rho_{G} \label{eq:gemm1} \\
\bm \nabla \wedge \mb E_{G}&=&-\frac{1}{c} \ppar{}{t}\left(\frac{\mb B_{G}}{2} \right) \label{eq:gemm2} \\
\bm \nabla \cdot  \left(\frac{\mb B_{G}}{2} \right)&=&0 \label{eq:gemm3} \\
\bm \nabla \wedge  \left(\frac{\mb B_{G}}{2} \right)&=&\frac{4\pi G}{c} \mb j_{G}+ \frac 1 c \ppar{\mb E_{G}}{t} \label{eq:gemm4}
\end{eqnarray}
Notice once again the factor $\frac 1 2$ near the gravito-magnetic field $\mb B_{G}$, with respect to the original Maxwell equations for the electromagnetic fields\footnote{We point out, again, that this is just a convention, introduced by Mashhoon \cite{mas93}, that can be used to exploit the standard results of electrodynamics to describe gravity in post-Newtonian linear approximation. Other conventions are used elsewhere \cite{ciufoliniwheeler,ruffi}. }. This is ultimately related to the fact that the linear approximation of GR involves a spin-2 field.

If the sources are stationary, the equation of motion (i.e. the spatial components of the geodesics) of a test mass $m$  moving with speed $\mb v$  in GEM fields $\mb E_{G}, \mb B_{G}$ turns out to be (see e.g. \cite{binind})
\beq
m\frac{\rmd {\mathbf v}}{\rmd t}=-m{\mathbf E_{G}}-2m \frac{{\mathbf v}}{c}\times {\mathbf B_{G}}, \label{eq:gemforce1}
\eeq
to lowest order in $v/c$. In the convention used, a test particle of inertial mass $m$ has gravito-electric charge $q_E=-m$ and gravito-magnetic charge $q_B=-2m$; the GEM Lorentz acceleration acting on a test particle is
\beq
\mb a=-{\mathbf E_{G}}-2 \frac{{\mathbf v}}{c}\times {\mathbf B_{G}} \label{eq:florentz}
\eeq

Notice that from the GM field $\mb B_{G}$, it is possible to obtain the precession rate of of a gyroscope (see e.g. \cite{mashh2}) in the GM field:
\beq
\bm \Omega_{G}=\mb B_{G}/c \label{eq:prOm}
\eeq
This effect is analogous to the magnetic precession of a dipole in a magnetic field, and its measurement was the goal of the GP-B mission \cite{GP.B}.

\section{GEM Fields of A Thin Ring}\label{sec:GEMf}

Let us describe the physical situation we are going to study. We consider an inertial frame, and a body of mass $M$ at rest in this reference frame; this body is slowly and uniformly rotating, and its angular momentum is $\mb S$. 

We suppose that the body is located at the origin of a Cartesian coordinate system $\{x,y,z\}$, and that its angular momentum is directed along the $z$ axis: $\mb S= S \mb u_{z}$. Hence, far from the source, the space-time is described by the metric
\beq
\mathrm{d} s^2= -c^2 \left(1-2\frac{\Phi}{c^2}\right)\rmd t^2 -\frac4c ({\mathbf A}\cdot \rmd {\mathbf x})\rmd t +
 \left(1+2\frac{\Phi}{c^2}\right)\Delta_{ij}\rmd x^i \rmd x^j\, \label{eq:weakfieldmetric1back}
\eeq
with (here and henceforth, for the sake of clarity, we drop the subscript ``$_{G}$'' to refer to GEM quantities)
\beq
\Phi= \frac{GM}{r}, \quad \mb A =  \frac{G}{c} \frac{\left(\mb S \wedge \mb x \right)}{r^{3}} \label{eq:PhiApert}
\eeq
where $r=|\mb x|=\sqrt{x^{2}+y^{2}+z^{2}}$, and these expressions are meaningful in the far region  of the field, where $\displaystyle r \gg \frac{GM}{c^{2}}$,  $\displaystyle r \gg \frac{S}{Mc}$. We see that, besides the usual Newtonian field deriving from $\Phi$, there is a gravito-magnetic contribution, deriving from $\mb A$, that has a dipolar behavior: $\displaystyle \mb B= \frac{G}{c} \left[\frac{3\left(\mb S \cdot \mb x \right)\mb x}{r^{5}}- \frac{\mb S}{r^{3}} \right]$. This is the term responsible for the Lense-Thirring effect (see e.g. \cite{ciufoliniwheeler,Iorio:2010rk}). 

Now, we suppose that the metric (\ref{eq:weakfieldmetric1back}) is perturbed by a ring of continuously distributed matter: the generic infinitesimal mass element $dm$ is orbiting a Keplerian ellipse around the central body; we know the total mass $m$ of the ring, and its angular momentum $\mb s$,  {which we assume to be \textit{constant}: in other words, we consider a stationary ring.} In our perturbative approach, we do suppose that $m\ll M$, $s \ll S$. Due to the presence of this ring, the GEM potentials are perturbed, so that $\Phi \rightarrow \Phi+\Delta \Phi$, $\mb A \rightarrow \mb A+\Delta \mb A$. Our purpose here is to calculate, by means of suitable power law expansions, the perturbations $\Delta \Phi, \Delta \mb A$ in terms of the mass and angular momentum  of the ring. In particular, we are interested in calculating these perturbations in the intermediate region between the central body and the ring; the same calculations can be carried out in the outer region of the system, i.e. away from the ring, by means of a similar multipole expansion, but here we are not interested in this case.

\subsection{Geometric Configuration}\label{ssec:conferring}

\begin{figure}[here]
\begin{center}
\includegraphics[scale=.40]{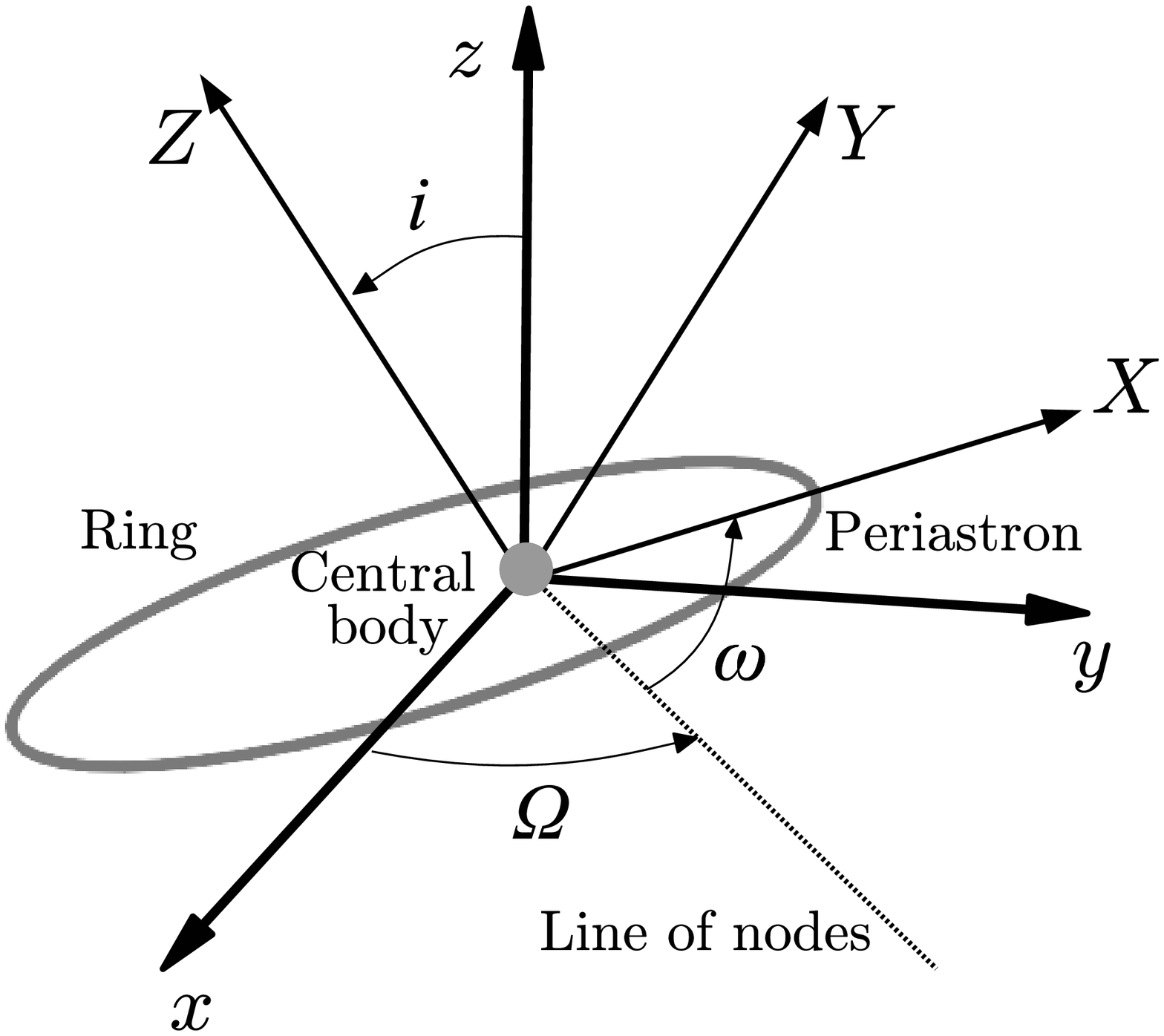}
\includegraphics[scale=.40]{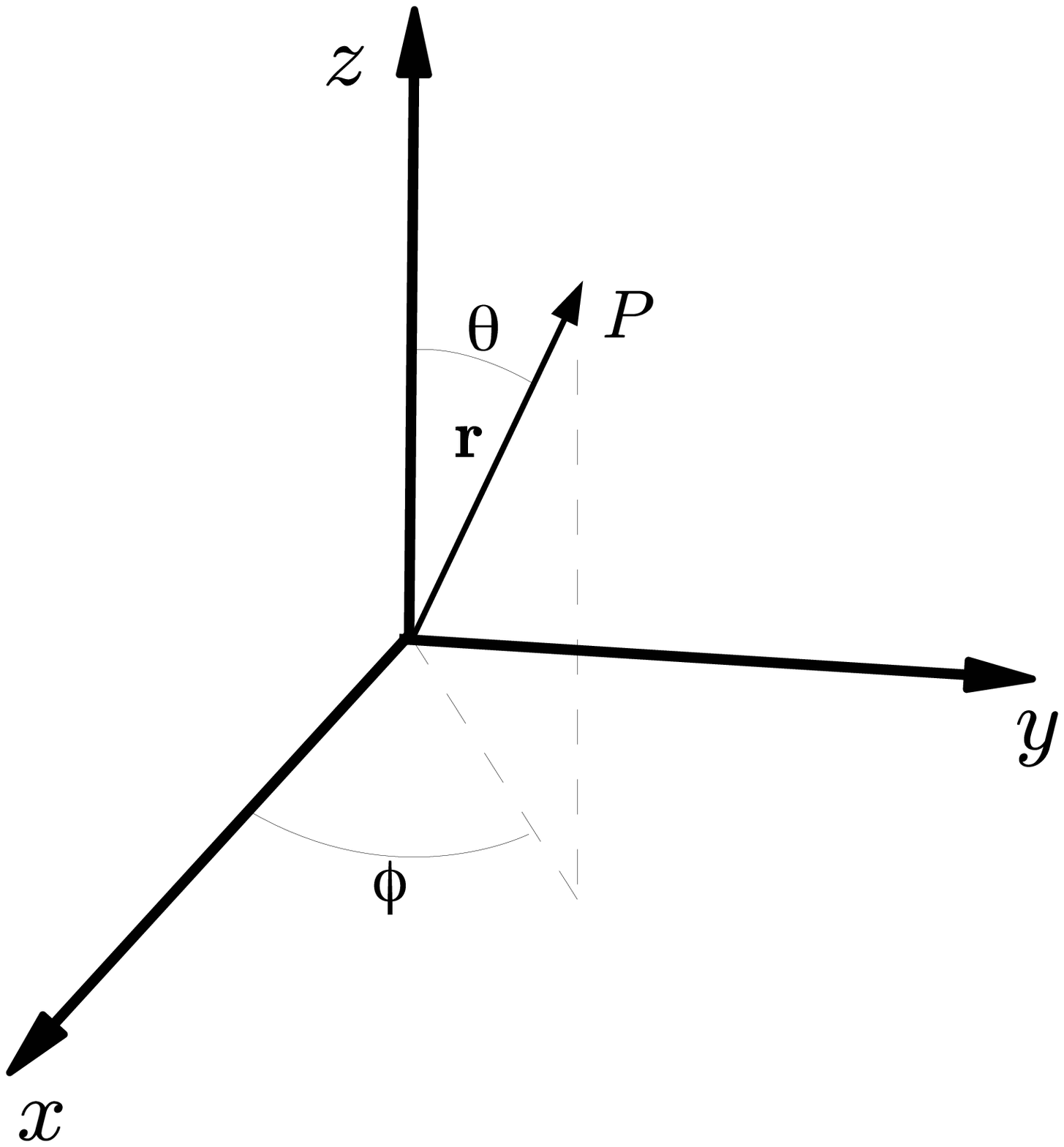}
\caption{Left: configuration of the ring. Right: The spherical coordinates $r,\theta,\phi$ are related to the Cartesian ones by $r=\sqrt{x^{2}+y^{2}+z^{2}}$,  $\phi=\arctan (y/x)$, $\theta=\arctan(\sqrt{x^{2}+y^{2}}/z$).} \label{fig:fig}
\end{center}
\end{figure}

We refer to Figure \ref{fig:fig}-left, to describe the geometry of the system we are considering.   For the sake of generality, we consider an arbitrary configuration of the ring that we describe in terms of the orbital elements. In the  inertial frame where the central body is at rest, we consider a primary Cartesian coordinate system $\{x,y,z\}$, with the corresponding unit vectors $\mb u_{x}, \mb u_{y}, \mb u_{z}$; then, we introduce another Cartesian coordinate system $\{X,Y,Z\}$, with the same origin, and unit vectors $\mb u_{X}, \mb u_{Y}, \mb u_{Z}$. The ring orbital plane is the $XY$ plane, and we denote with $\Omega$  the angle between the $x$ axis and the line of the nodes, while the angle between the $z$ and $Z$ axes is $i$. The ring periastron is along the $X$ axis, and we denote by $\omega$ the argument of the periastron, i.e. the angle between the line of nodes and the $X$ axis. 


The following relations hold between the unit vectors of the two Cartesian coordinate systems (see e.g. \cite{bertotti}):
\begin{eqnarray}
\mb u_{X} & = & \left(\cos \omega \cos \Omega-\sin \omega \cos i \sin \Omega \right) \mb u_{x}+\left(\cos \omega \sin \Omega+\sin \omega \cos i \cos \Omega \right) \mb u_{y}+\sin \omega \sin i  \, \mb u_{z} \nonumber \\
\mb u_{Y} & = & \left(-\sin \omega \cos \Omega-\cos \omega \cos i \sin \Omega \right) \mb u_{x}+\left(-\sin \omega \sin \Omega+\cos \omega \cos i \cos \Omega \right) \mb u_{y}+\cos \omega \sin i  \, \mb u_{z} \label{eq:trasxyzXYZ} \\
\mb u_{Z}&=& \sin i \sin \Omega \, \mb u_{x}-\sin i \cos \Omega \, \mb u_{y}+ \cos i \mb u_{z} \nonumber
\end{eqnarray}


Let $\mb X$ denote the position vector of a mass element $dm$ of the ring, with $R=|\mb X|$; then, the position of an arbitrary ring element $dm$ in the orbital plane is 
\beq
\mb X = R(f) \cos f \, \mb u_{X}+ R(f) \sin f \, \mb u_{Y} \label{eq:motionR}
\eeq
where the Keplerian ellipse, parameterized by the true anomaly $f$,  is written as
\beq
R(f)=\frac{a\left(1-e^{2}\right)}{1+e\cos f} \label{eq:keplerorbit1}
\eeq
in terms of the semi-major axis $a$ and eccentricity $e$.\\

In the orbital plane, it is useful to introduce the polar unit vectors\footnote{Usually defined by $\displaystyle \mb u_{R}= \cos f \, \mb u_{X}+ \sin f \, \mb u_{Y}, \, \, \, \mb u_{f} = -\sin f \, \mb u_{X}+\cos f \, \mb u_{Y}$.} $\mb u_{R}, \mb u_{f}$ so that $\mb X = R\, \mb u_{R}$ and the velocity vector is expressed by  $\displaystyle \mb v(f)= \frac{\sigma}{R} \left[g\left(f\right)\mb u_{R}+\mb u_{f} \right]$, where $\displaystyle g(f) \doteq \frac{e \sin f}{1+\cos f}$; $\sigma $ is the angular momentum per unit mass (see e.g. \cite{bertotti}),  {which is constant in the Keplerian motion.} Then, by straightforward algebra , it possibile to write: 
\beq
\mb v(f)= \frac{ \sigma}{R(f)} \left[g(f)\cos f-\sin f \right] \mb u_{X}+\left[g(f)\sin f+\cos f \right]\mb u_{Y}. \label{eq:motionv}
\eeq

We suppose that the  ring is homogenous, so that we can define its constant density $\displaystyle \lambda=\frac{m}{L}$, where $L$ is the length of the ring. In order to calculate the length of the ring, on using polar coordinates $\{R,f\}$ in the orbital plane, we may write the infinitesimal arc length in the form $\displaystyle \rmd L=\left(\rmd R^{2}+R^{2}\rmd f^{2}\right)^{1/2}$; then, using  Eq. (\ref{eq:keplerorbit1}), we obtain
\beq
\rmd L= R(f) \left[\frac{1+2e\cos f+e^{2}}{\left(1+e\cos f \right)^{2}} \right]^{1/2}\rmd f \label{eq:dL0}
\eeq
On integrating, and expanding in powers of the eccentricity to lowest approximation order in $e$, we obtain:
\beq
L=\int_{0}^{2\pi}R(f) \left[\frac{1+2e\cos f+e^{2}}{\left(1+e\cos f \right)^{2}} \right]^{1/2}\rmd f = 2\pi a \left( 1-\frac{e^{2}}{4} \right)\label{eq:dL}
\eeq

The above expressions will be used in what follows to calculate the GE and GM potentials.

\subsection{Gravito-electric perturbations}\label{ssec:GEME}

The contribution to the GE potential due to the ring is obtained by integrating the corresponding expression in (\ref{eq:solgemAi1}) over the whole ring. To this end, we substitute $ \lambda\, \rmd L=\rmd m$ for $\rho\, \rmd V$,  where $\displaystyle \lambda=\frac{m}{L}$, and we take into account (\ref{eq:dL0}) and (\ref{eq:dL}) to calculate $\lambda$. So, the integral we must evaluate is:
\beq
\Delta \Phi (\mb x) ={G}\int_{L}\frac{\lambda \rmd L}{|{\mathbf x}-{\mathbf X}|} \label{eq:deltaPhi1}
\eeq
In order to deal with the symmetries of the problem in a simpler way, we  express $\Delta \Phi$ (and, below, $\Delta \mb A$) as a function of the spherical coordinates $\{r,\theta,\phi\}$ (see e.g. Figure \ref{fig:fig}-right). Let $\mb x$ denote the position vector of an arbitrary point in the inertial frame, where we evaluate the GEM potentials and fields, with $r=|\mb x|$:  we express its components  as $\displaystyle
\mb x = r \, \sin \theta \cos \phi \,  \mb u_{x}+ r \,\sin \theta \sin \phi  \, \mb u_{y}+ r \, \cos \theta \,  \mb u_{z}$, while the components of the position vector $\mb X$ of the generic mass element $\rmd m$  along the ring are expressed by Eqs. (\ref{eq:trasxyzXYZ}) and (\ref{eq:motionR}).  As a consequence, we may write $\displaystyle |{\mathbf x}-{\mathbf X}|=\sqrt{r^{2}+R^{2}-2 \mb X \cdot \mb x}$, and expand in powers of $\displaystyle \frac{r}{R}$, since we are interested in the GEM potentials and fields in the intermediate zone between the central body and the ring, where $r<R$. 

Indeed, even if this procedure is straightforward and can be carried out without conceptual complications (see e.g. \cite{roy} and \cite{Iorio:2012tp}), the resulting expressions, even to lowest approximation order, are awkward for an arbitrary configuration of the ring.
For this reason, we are going to consider some simplifications  to emphasize the essential features of the problem.

To begin with, since $R$ is a function of the orbital parameters $a,e$, we perform an expansion in powers of $\displaystyle \frac{r}{a}$ and $e$. For an arbitrary configuration of the ring (i.e. $\Omega \neq 0, \, \omega \neq 0, \, i \neq 0$),  up to linear order in $\displaystyle \frac{r}{a}$ and $e$, we obtain the following expression for the GE potential of the ring
\begin{align}
\Delta \Phi &= \frac{Gm}{a} \left[ 1+\frac 1 2 e^{2}+ \frac 1 2 \frac r a e \left(\cos \omega \cos \Omega \sin \theta \cos \phi +\cos \omega \sin \Omega \sin \theta \sin \phi + \sin i \sin \omega \cos \theta \right. \right.+ \nonumber \\ & + \left. \left. \sin \omega \cos i \cos \Omega \sin \theta \sin \phi-\sin \omega \cos i \sin \Omega \cos \phi \right) \vphantom{\frac{1}{2}}  \right] \label{eq:deltaPhi1bis}
\end{align}


On using the  spherical unit vectors\footnote{The spherical unit vectors are expressed in terms of the Cartesian base by $\displaystyle \mb u_{r}=\sin\theta\cos\phi \mb u_{x}+\sin \theta \sin \phi \mb u_{y}+\cos \theta \mb u_{z}$, $\displaystyle \mb u_{\theta}=\cos\theta\cos\phi \mb u_{x}+\cos\theta \sin \phi \mb u_{y}-\sin \theta \mb u_{z}$, $\displaystyle \mb u_{\phi}=-\sin \phi \mb u_{x}+\cos \phi \mb u_{y}$.} $\mb u_{r},  \mb u_{\theta}, \mb u_{\phi}$ the corresponding GE field turns out to be 
\begin{align}
\Delta \mb E &= -\frac 1 2 \frac{Gme}{a^{2}} \left(\cos  \omega  \sin  \Omega
  \sin  \theta  \sin  {\it \phi}  -\sin
  \omega  \cos  i  \sin  \Omega
  \sin  \theta  \cos  {\it \phi}  +\sin
 i  \sin  \omega  \cos  \theta
  \right. \nonumber \\  & \left. + \sin  \omega  \cos  i  \cos
 \Omega  \sin  \theta  \sin  {\it \phi}
 +\cos  \omega  \cos  \Omega  \sin
  \theta  \cos  {\it \phi}   
 \right) \mathbf{u}_{r} \nonumber \\  & + \frac 1 2 \frac{Gme}{a^{2}} \left(  \sin  i
  \sin  \omega  \sin  \theta  -\cos
  \omega  \sin  \Omega  \cos  \theta
 \sin  {\it \phi}  -\cos  \omega  \cos
  \Omega   \theta  \cos  {\it \phi}
  \right. \nonumber \\ & \left. +\sin  \omega  \cos  i  \sin
  \Omega  \cos  \theta  \cos  {\it \phi}
  -\sin  \omega  \cos  i  \cos
  \Omega  \cos  \theta  \sin  {\it \phi}
  \right) \mathbf{u}_{\theta} \nonumber \\ &  - \frac 1 2 \frac{Gme}{a^{2}} \left(\cos  \omega  \sin  \Omega
  \cos  {\it \phi}  -\cos  \omega  \cos
  \Omega  \sin  {\it \phi}  +\sin  
\omega  \cos  i  \sin  \Omega  \sin
  {\it \phi}  +\sin  \omega  \cos  i
  \cos  \Omega  \cos  {\it \phi}  \right) \mathbf{u}_{\phi} \label{eq:deltaE1}
 \end{align}

If we suppose that the ring is in the $xy$ plane, i.e. in the plane orthogonal to the angular momentum of the central body (which is a natural  symmetry plane for the situation we are considering) and that the periastron is along the $x$ axis  (i.e. on setting $\Omega = 0, \, \omega = 0, \, i = 0$), we obtain  the following expressions for the GE potential  up to second order in $e$ and $r/a$: 
\begin{align}
\Delta \Phi&= {\frac {Gm}{a}}+ \frac 1 2 {\frac {Gm{e}^{2}}{a}}+\frac 1 2 \,{\frac {Gme\sin
  \theta  \cos  \phi  }{{a}^{}}} \frac r a + \left( 
\frac{3}{16} \,{\frac {Gm{e}^{2} \sin^{2} \theta  \cos^{2}  \phi    ^{}}{
{a}^{}}}-{\frac {69}{32}}\,{\frac {Gm{e}^{2}  \cos^{2} 
\theta    }{{a}^{}}} \right. \nonumber \\ & \left. +\frac 1 4\,{\frac {Gm}{{a}^{}}}+{
\frac {21}{32}}\,{\frac {Gm{e}^{2}}{{a}^{}}}-\frac 3 4\,{\frac {Gm 
\cos^{2}  \theta    }{{a}^{}}}\right) \frac{r^{2}}{a^{2}} \label{eq:deltaPhi2}
\end{align}
\normalsize


In  the $xy$ plane (where $\displaystyle \theta=\frac \pi 2$)  the potential (\ref{eq:deltaPhi2}) becomes
\beq
\Delta \Phi=+{\frac {Gm}{a}}+  \frac 1 2 \,{\frac {Gm{e}^{2}}{a}}+\frac 1 2 \,{\frac {Gme\cos
  \phi  r}{{a}^{2}}}+\frac{3}{16}\,{\frac {Gm{e}^{2}{r}^{2}
 \cos^{2} \phi  }{{a}^{3}}}+\frac 1 4 \,{\frac 
{Gm{r}^{2}}{{a}^{3}}}+{\frac {21}{32}}\,{\frac {Gm{e}^{2}{r}^{2}}{{a}^
{3}}} \label{eq:Phi0plane}
\eeq
Notice that the above expression becomes
\beq
\Delta \Phi={\frac {Gm}{a}}+\frac 1 4\,{\frac {Gm{r}^{2}}{{a}^{3}}} \label{eq:Phi0planecircle}
\eeq
for a circular ring, in agreement with \cite{Iorio:2012tp}. The  GE field in the  $xy$ plane turns out to be
\beq
\Delta \mb E = -\frac{Gm}{a^{2}}\left[\frac 1 2 e \cos \phi + \left( \frac 1 2 + \frac{21}{16}e^{2}+\frac 3 8  e^{2}\cos^{2} \phi  \right) \frac r a \right] \mb u_{r}+\frac{Gm}{a^{2}}e \sin \phi \left[\frac 1 2+\frac 3 8 e \cos \phi \frac r a  \right] \mb u_{\phi} \label{eqE0plane}
\eeq
On setting $e=0$, i.e. for a circular ring, we do obtain
\beq
\Delta \mb E= -\frac 1 2 \frac{Gmr}{a^{3}} \mb u_{r}  \label{eqE0planecircle}
\eeq



\subsection{Gravito-magnetic perturbations} \label{ssec:GEMB}

In order to write the GM perturbations, we proceed as in the previous Section, starting now from the expression (\ref{eq:solgemAi1}) of the GM potential. We must substitute $ j^{i}\rmd V \rightarrow \lambda v^{i} \rmd L$, where $v^{i}$ are the components of the velocity\footnote{It is useful to point out that, in this case, the motion of matter is Keplerian, so  the velocity is not constant along the ring: on the contrary, in a purely EM case, charge currents move with constant velocity along a wire.} of the matter elements along the ring, expressed by Eq. (\ref{eq:motionv}). We obtain
\beq
\Delta \mb A =\frac{2G}{c}\int_{L} \left\{ \frac{ \sigma}{R(f)} \left[g(f)\cos f-\sin f \right] \mb u_{X}+\left[g(f)\sin f+\cos f \right]\mb u_{Y} \right\} \frac{\lambda \rmd L}{|{\mathbf x}-{\mathbf X}|} \label{eq:deltaA0}
\eeq
We integrate the above expression along the ring, and then we perform an expansion in powers of $\displaystyle \frac r a $ and $e$. Notice that from  (\ref{eq:deltaA0}) we get the Cartesian components of $\Delta \mb A$, from which we may obtain the spherical components.

For an arbitrary configuration of the ring (i.e. $\Omega \neq 0, \, \omega \neq 0, \, i \neq 0$), up to linear order in $\displaystyle \frac{r}{a}$ and $e$, we obtain the following expressions for the spherical components of the GM potential:
\begin{align}
\Delta A_{r} & = \frac{Gs}{ca^{2}} \left(-2e \sin \omega \sin \Omega \sin \theta \sin\phi-2e \sin \omega \cos \Omega \sin \theta \cos \phi-2e \cos \omega \cos i \sin \Omega \sin \theta \cos \phi +\right. \nonumber \\ & \left.+2e\cos \omega \cos i \cos \Omega \sin \theta \sin \phi+2e \sin i \cos \omega \cos \theta   \right) \label{eq:deltaA1r}
\end{align}
\begin{align}
\Delta A_{\theta} & = \frac{Gs}{ca^{2}} \left[\vphantom{\frac r a }-2e \cos \theta \cos \phi \sin \omega \cos \Omega-2e \cos \theta \cos \phi \cos \omega \cos i \sin \Omega-2e\cos \theta \sin \phi \sin \omega \sin \Omega +\right. \nonumber \\ & \left.+ 2e \cos \theta \sin \phi \cos \omega \cos i \cos \Omega-2e \sin i \sin \theta \cos \omega - \left(-\sin \phi \sin i \sin \Omega-\cos \phi \sin i \cos \Omega \right)\frac r a \right] \label{eq:deltaA1theta}
\end{align}
\begin{align}
\Delta A_{\phi} & = \frac{Gs}{ca^{2}} \left[\vphantom{\frac 1 2 }2e\sin\phi\sin\omega\cos\Omega+2e\sin\phi\cos\omega\cos i \sin\Omega-2e \cos\phi\sin\omega\sin\Omega +\right. \\ & \left.+2e\cos\phi\cos\omega\cos i \cos \Omega+  \left(\cos i \sin\theta +\sin\phi\sin i \cos \theta \cos \Omega-\cos \phi \sin i \cos \theta \sin \Omega\right)\frac r a \vphantom{\frac 1 2 } \right] \label{eq:deltaA1phi}
\end{align}




The components of corresponsing GM field are
\begin{align}
\Delta B_{r} &= \frac{Gs}{ca^{3}} \left (2 \cos \theta \cos i + \frac{2\sin\phi\sin i \cos^{2}\theta \cos \Omega}{\sin\theta}-2\frac{\cos\phi\sin i \cos^{2} \theta \sin \Omega}{\sin\theta}+\right. \\ & \left.-2\frac{\sin \phi \sin i \cos \Omega}{\sin\theta}  +2\frac{\cos \phi \sin i \sin \Omega}{\sin \theta} \right) \label{eq:Brall}
\end{align}
\begin{align}
\Delta B_{\theta} &=\frac{Gs}{ca^{3}} \left(-2\cos i \sin \theta-2\sin \phi \sin i \cos \theta \cos \Omega+2 \cos \phi \sin i \cos \theta \sin \Omega \right) \label{eq:Bthetaall}
\end{align}
\begin{align}
\Delta B_{\phi} &=\frac{Gs}{ca^{3}} \left(-2\sin \phi \sin i \sin \Omega-2\cos \phi \sin i \cos \Omega \right) \label{eq:Bphiall}
\end{align}

As before, we consider the configuration with the ring in the $xy$ plane, with the additional condition that the the $x$ axis is directed toward the periastron. In this case, up to second order in $\displaystyle \frac r a$ and $e$, we get the following expression for the GM potential
\beq
\Delta A_{r}=\frac{Gs}{a^{2}} \left[ 2e \sin \theta \sin \phi+ \left(\frac 3 4 e^{2} \cos \phi \sin \phi \sin^{2} \theta \right) \frac{r}{a}+\left( \frac 1 4 e^{} \sin \theta \sin \phi -\frac 9 4 e \cos^{2}\theta \sin\theta \cos \phi\right)\frac{r^{2}}{a^{2}} \right] \label{eq:deltaAr11}
\eeq
\beq
\Delta A_{\theta}=\frac{Gs}{a^{2}} \left[ 2e \cos \theta \sin \phi+ \left(\frac 3 4 e^{2}  \cos \theta \sin \theta \cos \phi \sin \phi \right) \frac{r}{a}+\left(\frac 1 4 e \cos \theta \sin \phi -\frac 9 4 e \cos^{3}\theta \sin \phi\right)\frac{r^{2}}{a^{2}} \right] \label{eq:deltaAtheta11}
\eeq
\beq
\Delta A_{\phi}=\frac{Gs}{a^{2}} \left[ 2e \cos \phi+ \left(\sin \theta+\frac{21}{8} e^{2} \sin \theta+\frac 3 4 e^{2} \sin \theta \cos^{2}\phi\right) \frac{r}{a}+\left(\frac 7 4 e \cos \phi -\frac{15}{4}e \cos \phi \cos^{2} \theta \right)\frac{r^{2}}{a^{2}} \right] \label{eq:deltaAphi11}
\eeq
The corresponding GM field has the following components:
\beq
\Delta B_{r} = \frac{Gs}{ca^{3}} \left (2\cos \theta +6 e^{2} \cos \theta+9 e \cos \theta \sin \theta \cos \phi \frac r a \right) \label{eq:deltaBr11}
\eeq
\beq
\Delta B_{\theta}=-\frac{Gs}{ca^{3}} \left[2 \sin \theta+6 e^{2}\sin \theta+\left(5e\cos \phi-9e \cos \phi \cos^{2}\theta \right)\frac r a \right] \label{eq:deltaBtheta11}
\eeq
\beq
\Delta B_{\phi}=-\frac{4Gs}{ca^{3}} e \cos \theta \sin \phi \frac r a \label{eq:deltaBphi11}
\eeq

If we set $e=0$ (i.e. for a circular ring) we obtain the following expression for the GM field:


\beq
\Delta \mb B= \frac{Gs}{ca^{3}} \left(2\cos \theta \mb u_{r}-2\sin \theta \mb u_{\theta} \right) \label{eq:Bcircleorbitalplane}
\eeq
that is to say that $\Delta \mb B$ is directed along the $z$ axis, parallel to the angular momentum of the ring: $\displaystyle \mb B= \frac{2G}{ca^{3}} \mb s$. The above results can be compared to the magnetic field of a circular current loop (see e.g. \cite{jackson}, Ch. 5). In the $xy$ plane, the above expressions for the GM potential (\ref{eq:deltaAr11})-(\ref{eq:deltaAphi11}) and field (\ref{eq:deltaBr11})-(\ref{eq:deltaBphi11}) turn out to be
\begin{align}
\Delta \mb A &= \frac{Gs}{ca^{2}}\left(2e \sin \phi+\frac 3 4 e^{2}\cos \phi \sin \phi \frac r a+\frac 1 4 e \sin \phi \frac{r^{2}}{a^{2}} \right) \mb u_{r} \\ &+ \frac{Gs}{ca^{2}} \left[2e\cos \phi+\left(1+\frac{21}{8}e^{2}+\frac 3 4 e^{2} \cos^{2}\phi \right) \frac r a   +  \frac 7 4 e \cos \phi \frac{r^{2}}{a^{2}} \right] \mb u_{\phi} \label{eq:A0plane}
\end{align}

\beq
\Delta \mb B= -\frac{Gs}{ca^{3}} \left(2+6e^{2}+5\cos \phi e \frac r a \right)\mb u_{\theta} \label{eq:B0plane}
\eeq

Notice that for $e=0$  we get the following expressions for the GM potential
\beq
\Delta \mb A=  \frac{G}{c}\frac{s}{a^{2}} \frac r a \mb u_{\phi} \label{eq:A0planecircle}
\eeq
and field
\beq
\Delta \mb B= -\frac{2Gs}{ca^{3}} \mb u_{\theta} \label{eq:B0planecircle}
\eeq
or
\beq
\Delta \mb B= \frac{2G}{ca^{3}} \mb s \label{eq:B0planecirclebis}
\eeq

\section{Discussion}\label{sec:conc}

In the previous Section we have obtained the perturbations of the gravitational field due to presence of a matter ring, orbiting a central body. As we have seen, the expressions of the GEM potentials can be cumbersome when an arbitrary configuration of the ring is considered:  in particular, we point out that, as for the GE effects, our results in Section \ref{ssec:GEME} are  complementary to those of \cite{Iorio:2012tp} where only the case of a circular ring was taken into account (even though at higher approximation order).

Provided that both the mass $m$ and the angular momentum $\mb s$ of the ring are expected to be by orders of magnitude smaller than the mass $M$ and angular momentum $\mb S$ of the central body, it is anyhow important to evaluate the impact of these perturbations on observations, that can be used, in turn, to set constraints on the properties of the rotating ring. Actually, 
this approach was carried out in \cite{Iorio:2012tp} studying the GE perturbations, to infer bounds on the masses of annular distributions, by a comparison with the  data of the perihelia of the Solar System planets. It is interesting to  add some comments on the peculiarities of the GE perturbations: the behavior of the perturbation  $\propto r^{2}$  in Eq. (\ref{eq:Phi0planecircle})  is the same as that of the cosmological constant in the classical Schwarzschild-de Sitter solution (see e.g. \cite{Ruggiero:2007jr}). So, in principle, both rings of matter and the cosmological fluid produce similar effects, and this could be relevant in the determination of the cosmological constant from the analysis of the Solar System dynamics (see e.g. \cite{Iorio:2005vw}, \cite{Ruggiero:2006qv}). Notice, also, that the GE field inside a rotating  ring is not null, in contrast with what happens, for instance, in the case of a spherical rotating shell (see e.g. \cite{ciufolens}).



In what follows we focus on the GM perturbations to evaluate their impact  on some gravitational effects.
For the sake of simplicity, we choose a very simple situation: we suppose that a circular ring is in the $xy$ plane and we study its effect in the same plane. As a consequence, the GM potential and field are given by (\ref{eq:A0planecircle}) and (\ref{eq:B0planecirclebis}) (notice that in this case the GM field is constant).

To begin with, let us consider a spinning object at the center of the ring: according to Eq. (\ref{eq:prOm}), it precesses (with respect to an asymptotical inertial frame) with a precession rate given by
 {
\beq
\mb \Omega_{G}= \frac{2G}{c^{2}a^{3}} \label{eq:Omegaprec1}
\eeq}
This result is in agreement with (\cite{ashbygem}), where the issue of gravitational precession is discussed from a conceptual viewpoint. As a consequence, studying the precession of the object it is possible, in principle, to  infer information on the spin $\mb s$ of the ring: this can be applied, for instance, to the precession of the spin  $\mb S$ of the central body itself.

The presence of these GM perturbations can cause secular variations on the Keplerian orbits. According to Eq. (\ref{eq:florentz}), the GM extra acceleration due to the rotating ring is 
\beq
\Delta \mb a =  -2 \frac{{\tilde{\mathbf v}}}{c}\times {\Delta \mathbf B_{}} \label{eq:acclorentz1}
\eeq
where $\tilde{\mathbf v}$ is the speed of the body (considered as a test particle) that is supposed to move  in the orbital plane of the ring,  and ${\Delta \mathbf B_{}}$ is the GM field (\ref{eq:B0planecirclebis});  we notice that the perturbing acceleration is in the orbital plane and  is directed towards the center of the ring if the body and the ring rotate in the same direction.  {Notice that, due to the minus sign in Eq. (\ref{eq:florentz}),  two parallels mass currents repels each other, while anti-parallel currents attract each other: this is different from what happens in electromagnetism.}

Given the perturbing acceleration (\ref{eq:acclorentz1}), we can calculate its effects on planetary motions within the standard perturbative
schemes (see, e.g., \cite{roy}):  to this end, we consider the Gauss equations which enable us to study the perturbations of the Keplerian orbital elements.
 {We stress that, as we said before, the ring is stationary: in other words, the ring's matter motion is assumed to be constant during the particle's timescale. The case of a single mass element, revolving so quickly to be considered as a sort of continuos ring by the test particle, can be treated in the same way. Actually, this problem was faced and solved by Gauss, who used it to state his averaging theorem, according to which it is possible to  replace a perturbing body by an equivalent continuous
mass spread over its orbit: this does not change the secular effects, while it removes the periodic terms of the perturbation \cite{gauss}.}

We denote by  $\tilde a$, $\tilde e$,  $\tilde \omega$, $\tilde{\mathcal{M}}$  the semi-major axis, the eccentricity, the
argument of pericentre  and the mean anomaly, respectively.  We obtain non null secular variations only for the  argument of periastron
\beq
<\dot{\tilde{\omega}}>=  \frac{2 G s \tilde \sigma}{c^{2}\sqrt{1-\tilde e_{}^{2}}\tilde n \tilde{a}^{2}a^{3}} \label{eq:Deltaomega1}
\eeq
and for the  mean anomaly
\beq
<\dot{\tilde{\mathcal{M}}}>=  \tilde{n}+ \frac{6 G s \tilde \sigma}{c^{2}\tilde{n}{}\tilde{a}^{2}a^{3}} \label{eq:DeltaM1}
\eeq
where $\tilde \sigma$ is the angular momentum per unit mass of the planet, that can be written as $\tilde \sigma = \sqrt{GM\left(1-\tilde{e}^{2} \right) \tilde a}$, and $\tilde n$ is the mean motion. These relations can be used to make a comparison with the recent observations of the secular perihelion precessions  {\cite{fienga,pit1,pit2}}, to set constraints on the spin of matter rings in the Solar System.  {We point out that the mean anomaly cannot be used in these tests, because of the large uncertainty arising from the Keplerian mean motion.}  {As for the secular variation of the argument of periastron, the expression (\ref{eq:Deltaomega1}) can be simplified by substituting the angular momentum per unit mass of the test body $\tilde \sigma = \sqrt{GM\left(1-\tilde{e}^{2} \right) \tilde a}$, and the mean motion $\displaystyle \tilde n =\sqrt{\frac{GM}{\tilde a^{3}}}$. Accordingly, we obtain
\beq
<\dot{\tilde{\omega}}>=\frac{2Gs}{c^{2}a^{3}}
\eeq 
In other words,  $<\dot{\tilde{\omega}}>=\Omega_{G}$, where $\Omega_{G}$ is given by Eq. (\ref{eq:Omegaprec1}), so that it does not depend on the orbit of the test particle. Actually, we remember that in the simplified situation that we are considering (a circular ring and a  test body in the same plane)  the perturbations are determined by the constant GM field  (\ref{eq:B0planecirclebis}): it is then expected that these perturbations affect in the same way all test particles. As for their magnitude,we obtain
\beq
<\dot{\tilde{\omega}}>=s \, 2.9 \times 10^{-43} \left(\frac{1 \mathrm{AU}}{a} \right)^{3} \, \mathrm{mas} \ \mathrm{cty^{-1}}
\eeq
For instance, the spin $s$ of a hypothetical ring at $a=1$ AU can be constrained by using the perihelion of Venus, measured by \cite{fienga}:  $\Delta \omega_{\mathrm{Venus}}=0.2 \pm 1.5$ mas cty$^{-1}$; on neglecting Lense-Thirring and $J_{2}$ effects (see \cite{Iorio:2012tp}), we obtain $s \leq 5.9 \times 10^{42}$ kg m$^{2}$ s$^{-1}$. Similarly, we can constrain the  spin of the minor asteroids belt between Mars and Jupiter, by considering $a=2.8$ AU, and using the perihelion of Mars, measured by (\cite{fienga}) $\Delta \omega_{\mathrm{Mars}}=-0.04 \pm 0.15$ mas cty$^{-1}$; on neglecting the other precession effect, we obtain  $s \leq 8.3 \times 10^{42}$ kg m$^{2}$ s$^{-1}$. 
}

 {Indeed, the above estimates should be considered upper limits, useful to evaluate the order of  magnitude of the  effects. In fact, in actual physical situations, the GM perturbations due to the ring are present together with other effects, such as the Lense-Thirring  and the $J_{2}$ effects of the central body, the Newtonian or GE effects of the rings. If we focus on the effects of the ring, in order to separate the GE from the GM ones it is useful to emphasize that, even in this simplified model, the GE and GM perturbing fields are quite different. The GE field of a circular ring in its plane is given by (\ref{eqE0planecircle}): it produces an acceleration that is always directed to the center of the ring and that depends on the distance $r$ of the particle from the center. On the contrary, the GM field (\ref{eq:B0planecirclebis}) is constant: it produces an acceleration (\ref{eq:acclorentz1}) in the plane of motion and orthogonal to the velocity of the test particle; hence, both its direction and magnitude change along the orbit. Accordingly,   the GE secular variation of the argument of periastron depends on the orbit of the test particle (see also \cite{Iorio:2012tp})  while the GM variation is constant. These peculiarities could be exploited, in principle, in a systematic analysis of the secular perihelion precessions of the bodies of the Solar System, to constrain the mass and the spin of the  rings,  by taking into account the real geometric configurations. Similar analyses, of course, could be done in other astrophysical situations.}

The GM field of the rotating ring affects the propagation of light and provokes additional time delay in gravitational lensing. In \cite{ciufolens} the case of a rotating spherical shell was considered, and it was suggested that accurate measurements of the time delay could be used, in principle, to estimate the angular momentum of the celestial objects. We can rephrase here the procedure described in \cite{ciufolens} to evaluate the time delay for light rays propagating inside the rotating ring, in the orbital plane. If we write the GM potential (\ref{eq:A0planecircle}) in the form $\displaystyle \Delta \mb A= \frac{Gm}{ca} \bm \omega \times \mb r$, where $\bm \omega$ is the rotation rate of the ring (so that $s=mr^{2}\mathrm{\omega}$), we get the following expression for the GM time delay for a ray propagating along a straight line from the position $\mb r_{1}$ to $\mb r_{2}$: $\displaystyle \Delta t_{GM}=-\frac{2Gma}{c^{4}} \bm \omega \cdot \left(\hat{\mb r}_{1} \times \hat{\mb r}_{2} \right)$, where $\hat{\mb r}_{1}$, $\hat{\mb r}_{2}$ are the unit vectors. Even though the expression of the time delay is different from the case of a rotating shell, the order of magnitude is the same, and the same estimates apply.

\section{Conclusions}\label{sec:cconc}

The field equations of general relativity, in linear post-newtonian approximation, can be written in form of Maxwell equations, for the gravito-electric (or Newtonian field)  and gravito-magnetic fields (generated by the mass currents), in the gravito-electromagnetic analogy. In this paper  we have studied the GEM perturbations due to the presence of a matter ring, orbiting around a central body on a Keplerian orbit. In particular, we have considered a thin and homogenous ring and calculated its field in the intermediate zone between the central body and the ring. Motivated by the existence of astrophysical situations in which similar structures are present, we have obtained expressions of the GE and GM perturbations, up to linear order in $\displaystyle \frac{r}{a}$ and $e$, for arbitrary configurations of the ring, while  for the case of a ring in the equatorial plane of the central body, we have obtained relatively simple expressions accurate up to second order, thus generalizing and complementing some previous results.

In particular, we have focused on the GM perturbations and, considering for the sake of simplicity a rotating circular ring,   we have evaluated  the impact of these perturbations  on some gravitational effects, focusing on gyroscopes precession, Keplerian motion and time delay. 

We suggest that the combined analysis of the GE and GM effects could help to obtain, at least in principle, estimates on the mass \textit{and} on the angular momentum of  matter rings: to this end, we are going to undertake further   investigations  in a forthcoming paper.

\end{document}